\documentclass[11pt,a4paper]{article}
\pdfoutput=1
\usepackage{jcappub}
\usepackage{xcolor}

\newcommand{\dphi}{\delta\phi}

\newcommand{\mn}{\mu\nu}

\title{Cosmological instability of scalar-Gauss-Bonnet theories exhibiting scalarization}

\author[a]{T.~Anson,}
\author[a,b]{E.~Babichev,}
\author[a]{C.~Charmousis,}
\author[c]{S.~Ramazanov}

\affiliation[a]{Laboratoire de Physique Th\'eorique,
CNRS, Univ. Paris-Sud,\\
Universit\'e Paris-Saclay, F-91405 Orsay, France}
\affiliation[b]{Sorbonne Universit\'e, CNRS, UMR7095,
Institut d'Astrophysique de Paris,
${\mathcal{G}}{\mathbb{R}}\varepsilon{\mathbb{C}}{\mathcal{O}}$,\\
98bis boulevard Arago, F-75014 Paris, France}
\affiliation[c]{CEICO, Institute of Physics of the Czech Academy of Sciences, Na Slovance 1999/2, 182 21 Prague 8, Czech Republic}

\abstract{In a subclass of scalar-tensor theories, it has been shown that standard general relativity solutions of neutron stars and black holes with trivial scalar field profiles are unstable. 
Such an instability leads to solutions which are different from those of general relativity and have non-trivial scalar field profiles, in a process called scalarization.
In the present work we focus on scalarization due to a non-minimal coupling of the scalar field to the Gauss-Bonnet curvature invariant. 
The coupling acts as a tachyonic mass for the scalar mode, thus leading to the instability of general relativity solutions. 
We point out that a similar effect may occur for the scalar modes in a cosmological background, 
resulting in the instability of cosmological solutions. 
In particular, we show that a catastrophic instability develops during inflation within a period of time much shorter than the minimum required duration of inflation.
As a result, the standard cosmological dynamics is not recovered. This raises the question of the viability of scalar-Gauss-Bonnet theories exhibiting scalarization.
}

\begin{document}

\maketitle
\date{\today}

\section{Introduction}

Scalar-tensor
 theories with higher order derivatives~\cite{Horndeski:1974wa,Deffayet:2009wt,Deffayet:2009mn}  have attracted a lot of attention recently, as they represent testable modifications of general relativity (GR). A particular class of scalar-tensor theories involve the Gauss-Bonnet invariant, which is of second order in curvature. While per se this is a purely topological term in 4 dimensional metric theories, it becomes dynamical (and free of the Ostrogradsky ghost), when coupled to a scalar degree of freedom. The conditions imposed on the coupling function ensure that GR solutions with a trivial scalar field configuration always exist in this setup.
However, this scalar-tensor theory may have other solutions for black holes and stars, which are different from those of GR.  
The reason is that the Gauss-Bonnet term induces an effective mass for the scalar mode. When the effective mass is chosen to be tachyonic, GR solutions may become unstable in regions of strong curvature, while stable black hole or star solutions (if existent) acquire scalar hair absent in GR~\cite{Doneva:2017bvd,NS_Doneva,GB_Silva,Antoniou:2017acq,Antoniou:2017hxj}(See also \cite{Myung:2018iyq}). 
This phenomenon of co-existing solutions with the same symmetry but different scalar profiles has been dubbed scalarization. 
In fact scalarization with the Gauss-Bonnet term is quite similar to the well-known scalarization in ``standard'' scalar-tensor theories~\cite{gef93,gef_pulsars}, 
where the role of the effective mass is played by the Ricci scalar instead of the Gauss-Bonnet invariant. Interestingly, ``standard" scalarization operates only in the presence of matter, i.e., that composing neutron stars, while the Gauss-Bonnet term additionally allows for the scalarization of black holes. This is due to the fact that the Ricci scalar is zero in vacuum, unlike the Gauss-Bonnet invariant.

In this paper we point out that the term leading to the scalarization of compact objects may also be dangerous for the stability of cosmological solutions due to the same effect, namely the appearance of a tachyonic mass of the scalar mode\footnote{A similar study for the case of ``standard" scalarization has been done in~\cite{Anderson:2016aoi}.}. 
Indeed, as we will show in Sec.~\ref{sec:instability}, the mass squared of the scalar mode is positive in the case of a decelerating  Friedmann-Lema\^itre-Robertson-Walker (FLRW) Universe, while it is negative for an accelerating Universe. 
On the one hand, this means that the cosmological GR solution is safe at radiation and matter dominated epochs. That is, the presence of the Gauss-Bonnet term in the action does not destabilize the GR dynamics at those times. 
For the present-day accelerating phase, the effective mass of the scalar becomes tachyonic. Nevertheless, given the small Hubble rate the instability is very mild: 
it develops at time scales largely exceeding the age of the Universe. 
Therefore the solution stays very close to the GR cosmological solution for a long time.

On the other hand, the instability triggered by the scalar-Gauss-Bonnet coupling 
is disastrous during inflation, terminating the latter well before the `canonical' 60 e-folds elapse.
We demonstrate in Sec.~\ref{sec:inflation} that the primordial quantum fluctuations of the scalar field are quickly amplified to the point where they become dominant, 
thus destabilizing inflation. This suggests that the theory exhibiting scalarization via the coupling to the Gauss-Bonnet term may not be physically viable, and one needs to look for ultraviolet extensions or modifications of the theory. In Sec.~\ref{sec:discussion} we conclude with discussions on possible ways to avoid this catastrophic instability of the scalar mode during inflation.

\section{The model and scalarization of compact objects}
We consider the following action, which includes a coupling term $f(\phi)$ between a scalar field $\phi$ and the Gauss-Bonnet invariant $\hat{G}=R_{\mu\nu\sigma\alpha}R^{\mu\nu\sigma\alpha}-4R_{\mu\nu}R^{\mu\nu}+R^{2}$:
    \begin{equation}
    \label{action}
        S=\int \text{d}^{4}x\sqrt{-g}\left[\frac{M_P^2}{2}\left(R-2\Lambda\right)-\frac{1}{2}g^{\mu\nu}\partial_{\mu}\phi\partial_{\nu}\phi +f(\phi)\hat{G}\right] \; ,
    \end{equation}
where $\Lambda$ is the cosmological constant. The Euler-Lagrange equation for the scalar field reads:
    \begin{equation}
        \label{ELscalar}
        \square\phi+f'(\phi)\hat{G}=0 \; .
    \end{equation}
The variation of (\ref{action}) with respect to the metric yields modified Einstein equations:
\begin{equation}
\label{metriceq}
    M_P^2\left(G_{\mn}+ \Lambda g_{\mn}\right) = \nabla_\mu\phi\nabla_\nu\phi  - \frac{1}{2} g_{\mn}\nabla_\alpha\phi\nabla^\alpha\phi - 8P_{\mu\lambda\nu\alpha}\nabla^\alpha\left[f'(\phi)\nabla^\lambda\phi\right] \; .
\end{equation}
Here $P_{\alpha\beta\mu\nu}$ is the double dual of the Riemann tensor, defined using the antisymmetric Levi-Civita tensor $\varepsilon_{\mu\nu\alpha\beta}$ as:
	\begin{align}
		P_{\alpha\beta\mu\nu} & \equiv -\frac14 \varepsilon_{\alpha\beta\rho\sigma} R^{\rho\sigma\gamma\delta} \varepsilon_{\mu\nu\gamma\delta} \nonumber \\
			&= R_{\alpha\beta\mu\nu}+ g_{\alpha\nu} R_{\beta\mu} - g_{\alpha\mu} R_{\beta\nu} + g_{\beta\mu} R_{\alpha\nu}-g_{\beta\nu} R_{\alpha\mu} 
			+\frac12 \left( g_{\alpha\mu}g_{\beta\nu} - g_{\alpha\nu}g_{\beta\mu}\right) R\,.
	\end{align}
Under certain conditions on the coupling function $f(\phi)$, namely 
\begin{equation}
\label{cond}
f'(\phi_0)=0~\mbox{and}~f''(\phi_0)>0 
\end{equation}
for some constant $\phi_0$, such theories were shown to exhibit spontaneous scalarization around black holes and neutron stars~\cite{Doneva:2017bvd,NS_Doneva,GB_Silva,Antoniou:2017acq,Antoniou:2017hxj,Bakopoulos:2018nui}. This mechanism is similar to the one presented in \cite{gef93,gef_pulsars}, where the scalar field is 
coupled to the trace of the energy-momentum tensor and scalarization was shown to arise around neutron stars only. 
In the present case, scalarization results from the coupling of the scalar field to the Gauss-Bonnet invariant~\cite{Doneva:2017bvd,NS_Doneva,GB_Silva,Antoniou:2017acq,Antoniou:2017hxj,Bakopoulos:2018nui} and occurs for black holes as well.

The first condition on the coupling function ensures that $\phi=\phi_0$ passes through the field equations, in which case we recover the solution of general relativity (GR).
The second condition is crucial, as it implies that the scalar field acquires a negative effective mass, which is responsible for the appearance of scalar hair via a tachyonic instability. Indeed, one may study a scalar perturbation $\phi = \phi_0 +\delta\phi$ on a fixed Schwarzschild geometry, as was done in \cite{Doneva:2017bvd}. 
Notably, equations for the scalar and metric fluctuations are decoupled to the first order for constant $\phi_0$. The equation describing scalar perturbations is given by
\begin{equation}
        \left(\square+f''(\phi_{0})\hat{G}\right)\dphi=0 \; ,
\end{equation}
where the d'Alembert operator and the Gauss-Bonnet invariant are calculated with the Schwarzschild metric. One decomposes the scalar perturbation on the static and spherically symmetric background as 
\begin{equation}
\label{pert_scalar}
    \delta\phi = \frac{u(r)}{r}e^{-i\omega t}Y_{lm}(\theta,\varphi) \; ,
\end{equation}
where $Y_{lm}(\theta,\varphi)$ are the spherical harmonics. 
One can rewrite Eq.~(\ref{pert_scalar}) in the form of a Schr\"{o}dinger equation by introducing the tortoise coordinate $\text{d}r_*=\text{d}r\cdot(1-\frac{r_g}{r})^{-1}$, where $r_g$ is the Schwarzschild radius of the black hole:
\begin{equation}
\label{schr}
\frac{\text{d}^2u}{\text{d}r^2_*}+\omega^2 u = V_{\text{eff}}(r) u \; .
\end{equation}
For $l=0$ the effective potential $V_{\text{eff}}(r)$ reads
\begin{equation}
    V_{\text{eff}}(r) = \left(1-\frac{r_g}{r}\right)\left(\frac{r_g}{r^3}  - \frac{12 r_g^2}{r^6}f''(\phi_0 )   \right) \; .
\end{equation}
A sufficient condition on the effective potential for the existence of an unstable mode is \cite{potential}
\begin{equation}
    \int_{r_g}^{\infty}\text{d}r\frac{V_{\text{eff}}(r)}{1-\frac{r_g}{r}} < 0 \; .
\end{equation}
This condition, which may be satisfied if $f''(\phi_0)>0$, translates into
\begin{equation}
\label{cond3}
    r_g^2<\frac{24}{5}f''(\phi_0) \; .
\end{equation}
Thus the Schwarzschild solution becomes 
unstable for small enough masses, and one expects scalar hair to appear in that case.
Bearing in mind the possible redefinition $\phi\to \phi +\phi_0$, the function $f(\phi)$ can be expanded around $\phi_0=0$,
\begin{equation}
\label{coupling}
f(\phi)~=~\frac18\lambda^2\phi^{2}+\lambda^2\mathcal{O}\left(\phi^4/M_1^2\right) \; .
\end{equation}
Here $M_1$ is some scale normally taken to be of order of the Planck mass $M_P$, and the sign of the quadratic term is chosen so that the effective mass is tachyonic.
The value of $\lambda$ sets the upper bound of the mass of a black hole or a star at which scalarization may happen,
$\lambda \gtrsim r_g.$
For physically interesting objects, e.g., neutron stars, one easily finds 
\begin{equation}
	\lambda \sim \frac{M_\odot}{M_P^2}\sim 10^{19}\text{GeV}^{-1} \; .
\end{equation}
Different branches of scalarized solutions were shown to exist by a numerical analysis, for specific bands of $\lambda/r_g$.
Each branch may be labelled by an integer $n\in\mathbb{N}$ corresponding to the number of nodes of the radial scalar profile. It was shown in \cite{stability} that none of the branches are stable for a theory with a purely quadratic coupling function, $f(\phi)\propto \phi^2$. However, for the theory with $f(\phi)\propto (1-e^{-\phi^2/M_P^2})$ the fundamental branch $n=0$ is stable for a specific range of parameters. The same can be achieved by adding a quartic term to the purely quadratic coupling function \cite{quartic_1,quartic_2}. 

In the next sections we study the cosmological stability of scalar-Gauss-Bonnet theories with values of the coupling parameter $\lambda$ that lead to scalarized solutions around compact objects.

 \section{(In)stability of perturbations on FLRW background}
 \label{sec:instability}

In this section, we show that the tachyonic mass leading to the scalarization of black holes and stars 
is potentially dangerous on a cosmological background, and may result in a catastrophic instability of scalar modes.

Perturbing the EOM for the scalar field~(\ref{ELscalar}) one obtains
    \begin{equation}
        \label{scalar_pert}
        \left[\square+f''(\phi_{0})\hat{G}\right]\dphi=0 \; ,
    \end{equation}
 where  $\delta\phi$ is the perturbation of the scalar field.
Again, there are no terms involving $\delta g^{\mn}$ in the above equation since the scalar field is constant on the background. 
This equation defines an effective mass $m_{\text{eff}}$ for the scalar perturbation, where $m_{\text{eff}}^{2}=-f''(\phi_{0})\hat{G}$. 
Assuming the coupling~(\ref{coupling}), the effective mass squared can be written as
\begin{equation}
\label{meff}
m_{\text{eff}}^{2}=-\frac14 \lambda^2 \hat{G} \; .
\end{equation}
As we have discussed above, the case of real $\lambda$ corresponds to theories with scalarized stars and black holes.
For a flat FLRW background with the metric
    \begin{equation}
        \label{metric}
        \text{d}s^{2}=g_{\mu\nu}\text{d}x^{\mu}\text{d}x^{\nu}=-\text{d}t^{2}+a(t)^{2}\delta_{ij}\text{d}x^{i}\text{d}x^{j} \; ,
    \end{equation}
where $a(t)$ is the scale factor, the perturbation equation (\ref{scalar_pert}) takes the form
\begin{equation}
\label{perturbation}
\delta\ddot{\phi}+3H\delta\dot{\phi}-\frac{1}{a^{2}}\Delta\dphi +m_\text{eff}^2\delta\phi=0 \; .
\end{equation}
Here $H\equiv \dot a/a$ is the Hubble parameter, and we have assumed the coupling function of the form~(\ref{coupling}).
We now expand the scalar perturbation into Fourier modes,
$$
          \dphi \sim \int \text{d}\omega \text{d}^{3}{\bf k}\,\phi(\omega,{\bf k})e^{-i(wt-{\bf k}\cdot{\bf x})} \; ,
$$
to obtain the following dispersion relation:
\begin{equation}
            \omega^{2}= \frac{k^2}{a^2}+m_\text{eff}^2\label{dispersion} \; .
\end{equation}
In the above equation we have neglected the slow change of $\omega$ on the time scales shorter than $H^{-1}$.
Whether the mass of the scalar perturbations is real or tachyonic depends on the sign of $\hat{G}$ for the flat FLRW metric~(\ref{metric}),
    \begin{equation}
    \label{GBFriedmann}
        \Hat{G} =
        24H^{2}\frac{\ddot{a}}{a} \; .
    \end{equation}
Thus for a decelerating Universe, $\ddot{a}<0$, i.e., during radiation and matter dominated epochs, the mass of the scalar is real, $m_\text{eff}^2>0$, and no instability arises.
On the other hand, for an accelerated expansion, $\ddot{a}>0$, the mass (\ref{meff}) becomes tachyonic and the perturbations are unstable.
In particular, for a de Sitter solution with constant Hubble rate $H_{\text{dS}}$, one finds 
\begin{equation}
\label{meffdS}
	m_\text{eff}^2 = -6\lambda^2 H_{\text{dS}}^4 \; .
\end{equation}
The instability is extremely slow for the present-day acceleration of the Universe. 
Indeed, the ratio of the instability time $t_\text{inst}$ to the the age of the Universe $t_0$ is estimated as
\begin{equation}
	\label{instacceletaion}
	\frac{t_\text{inst}}{t_0} \sim \frac{H_0}{m_\text{eff}}\sim \frac{1}{\lambda H_0}\sim 10^{23} \; ,
\end{equation}
so that the instability is not noticeable. 
However, the same estimation for inflation with the scale $H_\text{inf}\sim 10^{13}$GeV and $N\sim 10^2$ e-foldings yields:
\begin{equation}
\label{instinfl}
	\frac{t_\text{inst}}{t_\text{inf}} \sim \frac{1}{N\lambda H_\text{inf}}\sim 10^{-34} \; .
\end{equation}
Thus the GR solution with $\phi=0$ has a very short instability time.
This suggests a catastrophic instability of the theory during the  inflation era, unless the initial value of the field $\phi$ 
is tuned to be extremely small. Indeed, as the scalar field grows as $\phi \sim \phi_1 e^{|m_\text{eff}| t}$ starting from some initial value $\phi_1$, one may choose $\phi_1$ arbitrarily close to 0, 
so that the instability does not develop during the time of inflation. In particular, if $\phi_1=\phi_0=0$, the field $\phi$ will stay on the top of the potential for an infinitely long time. 
However, even for the solution with $\phi_0=0$,  
quantum fluctuations of the scalar field rapidly grow and ultimately destroy the inflationary stage. 
This happens during a time which is much smaller than the duration of inflation, as we explicitly show in the next section.

\section{Catastrophic instability of the scalar field during inflation}
\label{sec:inflation}

In this section we show that the tachyonic instability inherent to the scalar field $\phi$ is inconsistent with the existence of inflation in the early Universe. We 
approximate inflation by an exact de Sitter expansion, which is a plausible assumption away from its final stages.   
From (\ref{metriceq}) one can find that in the presence of the field $\phi$ coupled to the Gauss--Bonnet term, the Friedmann equation is modified by the term
\begin{equation}
\label{relterms}
\rho_{\text{GB}} = - 6 \lambda^2 H^3 \phi \dot{\phi} \; .
\end{equation}
We require that the term~\eqref{relterms} be negligible compared to the inflaton energy density dominating the evolution of the Universe, i.e.,  
\begin{equation}
\label{condition}
\lambda^2 H^3 | \phi \dot{\phi}| \ll H^2 M^2_{P} \; .
\end{equation}
Our goal is to show that this condition gets violated quickly, namely as soon as inflation starts.
Even if the field $\phi$ is set at the top of the effective potential initially, $\phi=0$, its perturbations will cause a rapid instability. 
Perturbations of the field $\phi$ obey the equation (\ref{perturbation}) with the effective mass given by (\ref{meffdS}) with $H_{\text{dS}}=H$ being the Hubble scale during inflation.
We comment on deviations of the function $f(\phi)$ from 
a simple quadratic form at the end of this section. 

In the following we will primarily use the conformal time $\eta$, related to the cosmic time $t$ as
$$
\text{d}t = a \cdot\text{d}\eta \; .
$$ 
Conformal time is negative during inflation, and the scale factor verifies $aH = 1/|\eta|$.
The solution of the perturbation equation (\ref{perturbation}), assuming that the field $\phi$ is in the Bunch--Davies vacuum initially, is given by
\begin{equation}
\label{gen}
\phi ({\bf x}, \eta)=\int \frac{\text{d}^3{\bf k}}{(2\pi)^{3/2}} \frac{\sqrt{\pi}}{2} H |\eta|^{3/2} \left[ H^{(2)}_{\nu} (-k\eta) e^{-i{\bf k\cdot x}}A^{\dagger}_{{\bf k}}+ 
H^{(1)}_{\nu} (-k\eta) e^{i{\bf k\cdot x}}A_{\bf k} \right] \; ,
\end{equation}
where we have expanded $\phi$ into Fourier modes. Here $A_{{\bf k}}$ and $A^{\dagger}_{{\bf k}}$ are the annihilation and creation operators, respectively, and $H^{(i)}_{\nu} (-k\eta)$ are the Hankel functions of order 
\begin{equation}
\nonumber 
\nu=\sqrt{\frac{9}{4}-\frac{m_\text{eff}^2}{H^2}} \; .
\end{equation}
Note that the evolution of inflaton perturbations is described by $\nu \approx \frac{3}{2}$. In that case, perturbations get frozen as they exit the horizon. On the other hand, in the situation with $\nu \gg \frac{3}{2}$---the case of  our interest---perturbations grow fast beyond the horizon, as we will see shortly. This behaviour is due to the huge tachyonic mass acquired by the scalar field. From (\ref{gen}), the derivative $\frac{\partial}{\partial \eta} \phi $ is given by 
\begin{equation}
\nonumber
\frac{\partial}{\partial \eta} \phi ({\bf x}, \eta)=\int \frac{\text{d}^3{\bf k}}{(2\pi)^{3/2}} \cdot \frac{H\sqrt{\pi}}{2}  \cdot \left[ \left \{ -\frac{3}{2} |\eta|^{1/2} H^{(2)}_{\nu} (-k\eta) +|\eta|^{3/2} \frac{\partial}{\partial \eta} 
\left[H^{(2)}_{\nu} (-k\eta) \right] \right\} e^{-i{\bf k\cdot x}} A^{\dagger}_{{\bf k}}+h.c. \right] \; .
\end{equation}
The derivative of the Hankel function is expressed as follows:  
\begin{equation}
\nonumber 
\frac{\partial}{\partial z} H^{(1,2)}_{\nu} (z)=\frac{1}{2} \left[H^{(1,2)}_{\nu -1} (z)-H^{(1,2)}_{\nu+1} (z) \right] \; .
\end{equation}
We are working in the large $\nu$ regime. In this limit, the Hankel functions take the form: 
\begin{equation}
\label{asymptotics} 
H^{(1,2)}_{\nu} (-k \eta) \underset{\nu\rightarrow + \infty}{\sim} \mp i \cdot \left(\frac{2}{-k\eta} \right)^{\nu} \cdot \sqrt{\frac{2}{\pi \nu}} \cdot \left(\frac{\nu}{e} \right)^{\nu} \; ,
\end{equation}
(minus and plus signs correspond to the Hankel functions of the first and second kind, respectively). 
We see that it has a very sharp dependence on the order $\nu$. Hence, one can keep only Hankel functions 
with the largest $\nu$ when calculating the relevant correlation function: 
\begin{equation}
\label{intmain}
\langle \phi ({\bf x}, \eta) \frac{\partial}{\partial \eta} \phi ({\bf x}, \eta)  \rangle \simeq \int \frac{\text{d} k \cdot k ^3}{16\pi}  \cdot H^2    |\eta|^3 \cdot H^{(1)}_{\nu} (-k\eta)\cdot H^{(2)}_{\nu+1} (-k\eta)   \; ,
\end{equation}
where we made use of the commutation relation $[A_{\bf k}, A^{\dagger}_{{\bf k'}}]=\delta({\bf k}-{\bf k'})$ and integrated over the directions of the momenta ${\bf k}$.
As the lower limit of the above integral we choose some $k_{\text{min}}$, which is on-horizon at some moment $\eta_1$ {\it during} inflation, i.e., $k_{\text{min}} |\eta_1|=1$, but otherwise is arbitrary. We choose $|\eta_1|>|\eta|$, so that $\eta_1$ corresponds to the past with respect to $\eta$.
As for the upper limit  we take $k_{\text{max}}=\frac{1}{|\eta|}$. The result reads 
\begin{equation}
	\nonumber  \langle \phi ({\bf x}, \eta) \frac{\partial}{\partial \eta} \phi ({\bf x}, \eta) \rangle=\frac{H^2  }{8\pi^2 \nu |\eta|}  \cdot \left(\frac{2\nu }{e} \right)^{2\nu} \cdot \left[\left|\frac{\eta_1}{\eta} \right|^{2\nu-3}-1 \right]  \; .
	\end{equation}
Finally, we use that $\langle\phi\dot{\phi}\rangle = H|\eta|\langle \phi  \frac{\partial}{\partial \eta} \phi \rangle $ and implement condition~\eqref{condition} to obtain the ratio $\eta_1/\eta$ at which inflation gets violated: 
\begin{equation}
	\label{end}
	\left(\left|\frac{\eta_1}{\eta} \right|^{2\nu-3}-1 \right) \simeq \frac{8\pi^2\nu M^2_{P}}{\lambda^2H^4} \cdot \left[\frac{e}{2\nu} \right]^{2\nu} \; ,
	\end{equation}
or equivalently 
\begin{equation}
\nonumber 
\left| \frac{\eta_1}{\eta}\right| \simeq 1+{\cal O} \left(\frac{M^2_{P}}{\lambda^2 H^4} \cdot \left[\frac{e}{2\nu } \right]^{2\nu}\right) \; .
\end{equation}
We see that the modes with the momenta $k_1 \simeq \frac{1}{|\eta_1|}$ destabilize inflation immediately after exiting the horizon. Given an essentially arbitrary 
choice of $\eta_1$, we conclude that inflation cannot take place in this theory. 

One remark is in order here. Recall that our calculations imply the existence of modes which start from the Bunch--Davies vacuum and exit the horizon during inflation. This seemingly modest assumption requires some justification in the 
situation with exponentially large $\nu$. The asymptotic expansion of the Hankel functions for large positive arguments is given by 
\begin{equation}
\nonumber 
H_\nu^{(i)} (z) =\left(\frac{2}{\pi z} \right)^{1/2} \cdot e^{\pm i (z-\pi \nu/2-\pi/4)} \cdot \left(1+{\cal O} \left[\frac{\nu^2}{z}\right] \right) \; .
\end{equation}
The Bunch--Davies vacuum is defined in the limit $z=-k \eta \rightarrow +\infty$, when the second term in the brackets is irrelevant. In practice, however, 
inflation has a finite duration. Thus, the quantity $z$ is also large, but finite. Hence, the minimum value of momenta which are in the Bunch--Davies vacuum at the beginning of inflation is given by 
\begin{equation}
\label{minimal} 
k_{\text{min}} |\eta_i| \simeq \nu^2 \; .
\end{equation}
It is possible, in principle, that the modes with these large momenta never exit the horizon during inflation, formally invalidating our analysis. 
Let us show that this is not the case, unless the duration of inflation is tuned to its minimum value. 
By integrating  Eq.~\eqref{intmain} from $k_{\text{min}}$ to infinity (i.e. $|\eta| \ll 1$), we get
\begin{equation}
\nonumber 
\langle \phi \frac{\partial}{\partial \eta} \phi \rangle =\frac{H^2}{8\pi^2 \nu |\eta|} \cdot \left(\frac{1}{k_{\text{min}} |\eta|} \right)^{2\nu-3} \cdot \left(\frac{2\nu}{e} \right)^{2\nu} \; .
\end{equation}
We now apply the condition Eq.~\eqref{condition}, leading to
\begin{equation}
\nonumber 
k_{\text{min}} |\eta| \gg \nu \; .
\end{equation}
Combining the latter with Eq.~\eqref{minimal}, we obtain
\begin{equation}
\nonumber 
\left|\frac{\eta_i}{\eta} \right| \ll \nu \; ,
\end{equation}
which should hold until the end of inflation $\eta = \eta_f$. Unless the duration of inflation is close to $60$ e-foldings, the ratio $|\eta_i/\eta_f|$ is very large, even compared 
to the huge value $\nu \simeq 10^{32}$ used for scalarization of the stars. Hence, the modes, which are in the Bunch--Davies initially and exit 
the horizon during inflation, {\it alone} destabilize inflation already $75$ e-folds after it starts. 

Let us also comment on a possible stabilization of the scalar field due to the presence of the quartic corrections, 
since one should include those anyway to make the scalarized branch of compact objects physically viable~\cite{quartic_1,quartic_2}. 
It is not difficult to see, however, that the quartic terms cannot stabilize the field $\phi$ and prevent the catastrophic instability, because destabilization occurs at tiny values of $\phi$, for which the approximation $f(\phi) \propto \phi^2$ still holds. 
Indeed, from Eq.~\eqref{gen} we obtain
\begin{equation}
\nonumber 
\langle \phi^2 ({\bf x}, \eta) \rangle =\int \frac{\text{d}k \cdot k^2}{8\pi}  \cdot H^2  |\eta|^3 \cdot |H^{(1)}_{\nu} (-k \eta)|^2 \; . 
\end{equation}
Using Eq.~\eqref{asymptotics} and integrating over the modes in the range $\left(\frac{1}{|\eta_1|}, \frac{1}{|\eta|} \right)$, we get
\begin{equation}
	\label{unstable}
	\langle \phi^2 ({\bf x}, \eta) \rangle = \frac{H^2}{8\pi^2 \nu^2}  \cdot \left(\frac{2\nu }{e} \right)^{2\nu}\cdot \left(\left|\frac{\eta_1}{\eta} \right|^{2\nu-3}-1 \right) \; .
	\end{equation}
Finally, substituting Eq.~\eqref{end} into Eq.~\eqref{unstable}, we obtain the typical value of the scalar when inflation is violated:
\begin{equation}
\nonumber 
\langle \phi^2 ({\bf x}, \eta) \rangle  \simeq \frac{M^2_{P}}{\lambda^3 H^3} \; ,
\end{equation} 
where we used that $\nu \simeq \lambda H$. Given that $\lambda H \simeq 10^{32}$, we conclude that the instability develops at the field values $\sqrt{\langle \phi^2 ({\bf x}, \eta) \rangle}\sim 10^{-48}M_{P} \lll M_{P}$ in the theory exhibiting scalarization, i.e., well before the quartic term starts to act.

\section{Discussion and conclusions}
\label{sec:discussion}
In this paper we studied the stability of cosmological solutions in theories exhibiting scalarization of black holes and neutron stars due to the coupling of the scalar to the Gauss-Bonnet invariant. 
The scalarization of compact objects happens because the scalar acquires a tachyonic mass and its perturbations destabilize the GR solutions. 
It is natural to question whether an analogous destabilization arises on FLRW backgrounds. This has been the subject of our study.

As we have shown in Sec.~\ref{sec:instability}, stability of the scalar perturbations on GR cosmological solutions depends on the sign of the acceleration $\ddot{a}$, see Eqs.~(\ref{meff}) and (\ref{GBFriedmann}). 
For a decelerating Universe, the mass in the equation for the scalar perturbation is real, therefore no instability arises for the GR branch.
However, the mass becomes tachyonic for an accelerated expansion, and one expects the GR cosmological solution to become unstable in that case. 
It turns out that this instability is very slow with respect to the current acceleration, with the time of instability being much larger than the age of the Universe, see Eq.~(\ref{instacceletaion}). 

On the contrary, the scalar-Gauss-Bonnet coupling leads to a catastrophic instability during the inflationary epoch, due to its huge effective tachyonic mass. 
The characteristic instability time, at which the amplitude of a classical solution increases by the factor $e$, 
is given in Eq.~(\ref{instinfl}). This is a tiny number compared to the time of inflation.
One should take into account that classically the initial conditions of the scalar field can be tuned such that the field stays on top of the potential for an infinite time, at least in principle.
However, as we explicitly show in Sec.~\ref{sec:inflation}, quantum fluctuations of the scalar field are strong enough to trigger this instability, 
making any conventional inflationary scenario impossible in the theories exhibiting scalarization under study (\ref{action}). 

Throughout this study we have assumed that the scalarization field $\phi$ is {\it{not}} the inflaton field. However, the Gauss-Bonnet scalar is an ultraviolet correction which naturally modifies gravity at early times. Indeed inflation models where the Gauss-Bonnet scalar is coupled to the inflaton field according to our action (\ref{action}) have been studied in the literature (see for instance \cite{Kanti:2015pda, Kanti:2015dra, Hikmawan:2015rze}). However, in these studies, the effective sign of the coupling function $f$ in (\ref{action}) is crucially required to be of the opposite sign to the one required for scalarization.

It would be interesting to find ways to stabilize the theories exhibiting scalarization during inflation. For instance, one may try to solve the problem by adding a coupling to the inflaton $\chi$, e.g., $g \chi^2\phi^2$, where $g$ is the coupling constant. In this case the inflaton expectation value serves as a stabilizing effective mass, which vanishes as inflation ends. The problem here is that the coupling constant $g$ should be huge in this approach. Indeed, to balance the large value of $m^2_{\text{eff}}$ given by Eq.~(\ref{meffdS}), 
one needs to assume that 
\begin{equation}
\nonumber 
g \gtrsim \frac{6\lambda^2 H^{4}}{\chi^2} \simeq 10^{53} 
\end{equation} 
for $\chi \simeq M_P$, $\lambda H \simeq 10^{32}$, and $H \simeq 10^{13}$~\mbox{GeV}. Thus, the theory is deeply in the strong coupling regime, where no trustworthy predictions can be made. On the other hand, as it has been shown in \cite{Anson:2019ebp}, 
this coupling does fix a similar problem in the original model of scalarization 
by Damour and Esposito-Far\`ese \cite{gef93,gef_pulsars} without putting the theory in the strong coupling regime.

Another idea would be to add an extra coupling of the scalar to higher powers of curvature, so that it becomes dominant during inflation.
The extra term should be higher than the second order in the curvature,
otherwise one risks to spoil scalarization triggered by the coupling to the Gauss--Bonnet invariant, which is of the second order in the curvature. 
For example, the coupling $\sim \phi^2 R^4$ with the appropriate sign stabilizes the scalar field in the high-curvature regime. 
In this case, similarly to $f(R)$ gravity, an extra scalar degree of freedom is effectively introduced\footnote{We thank Gilles Esposito-Far\`ese for pointing this out.}, which is coupled to $\phi$. 
However, the coupling between $\phi$ and the scalar from the gravitational sector must be huge in order to balance the term involving the Gauss--Bonnet invariant. As we have discussed above, the introduction of a strong coupling is not a viable solution.

Yet another possible way to fix the catastrophic instability during inflation is to introduce the quartic terms in the scalar, $\sim \phi^4 \Hat{G}$. 
It has been shown that such terms help to stabilize the scalarized solutions of compact objects, see, e.g., Refs.~\cite{quartic_1,quartic_2}. 
However, as we have seen in Sec.~\ref{sec:inflation}, destabilization of inflation happens already for very small $\phi$, where the quartic term does not play any role. Adding a large coupling $\tilde{g}$, i.e., writing $\tilde{g} \phi^4 \Hat{G}$
 does not improve the situation for the following reason. We presume that the solutions for the scalarized compact objects will be indistinguishable from GR ones, since the scalar field will have values extremely close to zero due to the stabilizing quartic term.
A very similar idea has been suggested recently in \cite{Macedo:2019sem}, 
where it has been argued that a quartic term $\sim g\phi^4$, added to the action to stabilize 
the scalarized solution of compact objects, may also solve the problem of instability during inflation. 
In \cite{Macedo:2019sem} it has been suggested that the scalar field is in the minimum of the effective potential during 
inflation. It is easy to see, however, that this mechanism cannot work either. First, 
the minimum should be set at a value of order of $10^{64} M_P$ (in dimensionful units), 
which is highly unnatural. Moreover, as inflation ends, the structure of the effective potential 
changes because the scalar-Gauss-Bonnet coupling changes sign, 
and the scalar tends to the new minimum located at zero. 
Carrying a tremendous initial potential energy density estimated as $\sim 10^{180}M_P^4$, the rolling scalar field again 
causes a gross modification of the standard cosmological picture.

\acknowledgments{We thank Gilles Esposito-Far\`ese for interesting discussions. 
T.A., E.B. and C.C. acknowledge support from the research program
PRC CNRS/RFBR (2018--2020) n\textsuperscript{o}1985 ``Gravit\'e modifi\'ee et trous noirs: signatures exp\'erimentales et mod\`eles consistants''. 
S.R. is supported by the European Regional Development Fund and the Czech Ministry of Education, Youth and Sports through the Project CoGraDSCZ.02.1.01/0.0/0.0/15 003/0000437.


\end{document}